\documentclass[letters,usenatbib]{mnras}
\usepackage{graphicx,subfigure,amsmath, amsfonts, amssymb,footnote,color,epstopdf}
\usepackage{apjfonts}

\newcommand\phn{\phantom{0}}%
\title{Direct Detection of Quasar Feedback Via the Sunyaev-Zeldovich Effect}

\author[M. Lacy et al.]{Mark Lacy$^1$, Brian Mason$^1$, Craig Sarazin$^2$, Suchetana Chatterjee$^3$, Kristina Nyland$^1$, 
\newauthor{Amy Kimball$^4$, Graca Rocha$^5$, Barnaby Rowe$^6$, Jason Surace$^7$}
\\
$^{1}$National Radio Astronomy Observatory, 520 Edgemont Road, Charlottesville, VA 22903, USA\\
$^{2}$Department of Astronomy, University of Virginia, 530 McCormick Rd., Charlottesville, VA 22904, USA \\
$^{3}$Department of Physics, Presidency University,  Kolkata, 700073, West Bengal, India\\
$^{4}$National Radio Astronomy Observatory, 1003 Lopezville Road, Socorro, NM 87801, USA\\
$^{5}$Jet Propulsion Laboratory, California Institute of Technology, 4800 Oak Grove Drive, Pasadena, CA 91109, USA\\
$^{6}$Department of Physics and Astronomy, University College London, Gower Street, London WC1E 6BT, UK\\
$^{7}$Infrared Processing \& Analysis Center, MS 100-22, California Institute of Technology, Pasadena, CA, 91125, USA
}

\date{Last updated ; in original form }
\pubyear{2018}

\begin{document}
\label{firstpage}
\pagerange{\pageref{firstpage}--\pageref{lastpage}}
\maketitle



\begin{abstract}
The nature and energetics of feedback from thermal winds in quasars can be constrained via observations of the Sunyaev-Zeldovich Effect (SZE) induced by the bubble of thermal plasma blown into the intergalactic medium by the quasar wind. In this letter, we present evidence that we have made the first detection of such a bubble, associated with the hyperluminous quasar HE~0515-4414. The SZE detection is corroborated by the presence of extended emission line gas at the same position angle as the wind. Our detection appears on only one side of the quasar, consistent with the SZE signal arising from a combination of thermal and kinetic contributions. Estimates of the energy in the wind allow us to constrain the wind luminosity to the lower end of theoretical predictions, $\sim 0.01$\% of the bolometric luminosity of the quasar. However, the age we estimate for the bubble, $\sim 0.1$~Gyr, and the long cooling time, $\sim 0.6$~Gyr, means that such bubbles may be effective at providing feedback between bursts of quasar activity. 
\end{abstract}






\section{Introduction} \label{sec:intro}

Outflows from Active Galactic Nuclei (AGN) and starbursts are one of the major sources of feedback in galaxy evolution. Their interaction 
with the interstellar medium (ISM) of the galaxy can inject turbulence, dissociate molecular gas, or even drive the gas out of the galaxy completely (e.g.\ Silk \& Rees 1998; Bower et al.\ 2006; Hopkins et al.\ 2006; Croton et al.\ 2006; Richardson et al.\ 2016). AGN feedback is usually 
classified as one of two modes. A powerful quasar outburst can launch hot winds on short timescales in the ``quasar" or ``radiative'' mode. On longer timescales, lower power outflows associated with jets of relativistic plasma can provide ``radio'' or ``kinetic'' mode feedback (Fabian 2012). The relative roles of these two modes in providing feedback to stifle star formation in the host galaxy is still unclear. While the quasar mode is dramatic, with high velocity outflows seen in ionized gas (e.g.\ Harrison et al.\ 2014, Liu et al.\ 2014), much of the dense molecular gas entrained in the wind fails to reach escape velocity and will ultimately fall back to form stars  (Alatalo 2015; Emonts et al.\ 2017). In contrast, 
radio jets are effective at blowing bubbles of plasma into the intergalactic medium (IGM) on the ISM of the host galaxy is subtle,  limited to inducing 
turbulence in the ISM (e.g.\ Alatalo et al.\ 2015; Lanz et al.\ 2016; McNamara et al.\ 2016). 

An important step towards understanding the nature of AGN feedback on galaxies is estimating the energy of a wind or outflow in the quasar mode. In many models, the energetically-dominant phase in the outflowing gas is hot ($\sim 10^7$~K) with low density (e.g., Faucher-Gigu\`{e}re \& Quataert 2012; Zubivas \& King 2012). 
There have been a few detections of AGN outflows in X-rays (e.g., Greene et al.\ 2014; Sartori et al. 2016; Lansbury et al.\ 2018), but the tenuous nature of the hot phase gas, combined with the presence of a bright point source AGN, makes these energetics estimates challenging (Powell et al.\ 2018).  An alternative way to detect the hot gas phase is via the Sunyaev-Zeldovich Effect (SZE; Sunyaev \& Zeldovich 1972). The SZE is the spectral distortion of the cosmic microwave background (CMB) radiation due to the inverse Compton scattering of the CMB photons by the energetic electrons present along its line-of-sight. 

A thermal wind from an AGN produces a bubble of hot gas that is overpressured compared to the surrounding IGM. Thus, as first suggested by Natarajan \& Sigurdsson (1999), it should be possible to detect the SZE towards winds from powerful quasars or highly-luminous starbursts (e.g.\ Chatterjee \& Kosowsky 2007; Chatterjee et al.\ 2008; Scannapieco et al. 2008; Rowe \& Silk 2011; hereafter RS11). 
If these bubbles persist in the IGM for long periods ($\sim1$~Gyr) without cooling, they could also 
act as agents of feedback in much the same way as the plasma bubbles blown by radio jets in the radio mode.

Statistical studies using stacked data from single-dish telescopes have detected significant signals from quasar hosts (Chatterjee et al. 2010; Ruan et al. 2015; Crichton et al.\ 2016; Verdier et al.\ 2016), but it is unclear whether these results are affected by contamination of the SZE by either the intragroup or intracluster medium around the quasar, or from star formation in the quasar host (Cen \& Safarzadeh 2015; Soergel et al.\ 2017; Dutta Chowdhury \& Chatterjee 2017). Another approach has been to stack data on quiescent elliptical galaxies, where contamination is less of an issue, and fossil winds from prior AGN activity may persist (e.g., Spacek et al.\ 2016). 

Attempts to directly detect the SZE from thermal winds can be made with interferometers such as the Atacama Large Millimeter Array (ALMA; e.g. Chatterjee \& Kosowsky 2007; RS11). The predicted size scale of the signal, $\sim 10-100$~kpc, corresponding to $\sim 1-10$~arcsec at $z\stackrel{>}{_{\sim}}1$, is well matched to the 
angular resolution of ALMA in the most compact configurations. In this paper, we describe our attempt to use ALMA to 
directly detect the SZE from a quasar 
wind. We selected the most luminous radio-quiet quasar we could find in the literature that had good visibility to ALMA, HE~0515-4414 ($z=1.71$; Reimers et al.\ 1998). We assume a $\Lambda$ CDM cosmology
with
$H_0=70$ km s$^{-1}$ Mpc$^{-1}$,
$\Omega_{\rm M}=0.3$ and
$\Omega_{\Lambda}=0.7$.

\section{Observations and analysis}

In this Section we describe the ALMA observations of HE~0515-4415. We also discuss near-infrared (NIR) observations we obtained with the {\em Spitzer Space Telescope} and Gemini telescope, and archival {\em Hubble Space Telescope (HST)} data.


\subsection{ALMA}\label{sec:alma}

The peak intensity of the thermal SZE decrement is seen at
$\approx 130$ GHz (Sunyaev \& Zeldovich 1972), which lies in ALMA band-4.
Our observations were of a single pointing centered on the quasar using 
four 2~GHz basebands centered at 133, 135, 145, and 147~GHz. Two scheduling blocks were made: one  executed once in a relatively large configuration 
(delivering $\theta_{\rm FWHM}$ $\approx$0\farcs7)
to measure the point source contribution, and one executed 14 times in a compact configuration 
(delivering $\theta_{\rm FWHM}$ $\approx$2\farcs7) to maximize our response to the SZE. A standard calibration strategy was used resulting in an amplitude calibration accuracy of $\approx 5$\%\footnote{https://almascience.nrao.edu/documents-and-tools/cycle4/alma-technical-handbook. Calibration of the data was performed with the ALMA pipeline.}


In order to establish the presence or otherwise of the SZE  from the quasar host it was first necessary to 
subtract the emission from the quasar itself, and also the three other bright ($>20\, \mu$Jy) sources in the field (A1, A3, and A5; Figure~\ref{fig:he0515_multiwavelength}). Fortunately, all sources were unresolved in the smaller configuration data, and the subtraction of point source models in the $uv$-plane  
was adequate to remove them, with the possible exception of a small amount of diffuse extended emission associated with the brightest source, A5.

\begin{table}
\caption{Point source emission model subtracted from the ALMA data}
\label{tab:model}
\begin{tabular}{lcccc}
Source & RA & Dec & Flux density \\
             &.     &        & ($\mu$Jy)\\\hline\hline
A1&05:17:07.767&-44:10:43.87&{\phn41.7} \\
A3&05:17:08.483&-44:10:51.22&{\phn21.8}\\
A5&05:17:09.455 & -44:10:52.39 & 188.2\\
QSO&05:17:07.614&-44:10:55.64&{\phn66.2} \\\hline

\end{tabular}
\end{table}

No clear signal was readily apparent in the naturally-weighted map (which has the highest point source sensitivity, with an RMS of $\sigma =$ 
3.5 $\mu$Jy~beam$^{-1}$ 
and a synthesized beam of $\theta_{\rm FWHM} = $ 3\farcs2$\times$2\farcs3 at a position angle of PA = 106$^{\circ}$), so tapering was applied to improve the surface brightness sensitivity of the image. The taper with a beam best matched to the size of the signal 
produced an image with $\sigma =$ 6.6 $\mu$Jy~beam$^{-1}$ and $\theta_{\rm FWHM} = $ 6\farcs7$\times$6\farcs4 at PA $= 110^{\circ}$, and recovers an apparent dip of flux density of $-23.1\,\mu$Jy, significant at the 3.5$\sigma$ level, to the SW of the quasar position. (After a 12\% correction for the response of the primary beam, this peak value of the decrement becomes $-25.9 \, \mu$Jy.) Figure~\ref{fig:he0515_multiwavelength} shows both the ALMA naturally-weighted image (prior to source subtraction) and tapered image (after source subtraction) as contours superposed on greyscales of the near-infrared imaging from {\em Spitzer}, Gemini, and {\em HST} described below. 

\subsection{Optical and Infrared}

{\em Spitzer} observations were obtained in a DDT program (PID 13221). 
These observations used 40 dithers from the standard large random dither pattern and 30s frametimes to achieve a 5$\sigma$ depth of AB$\approx 23.1$ in the 
$3.6$ and $4.5\,\mu$m bands. 
The standard pipeline products were used. 


Gemini fast turnaround observations were obtained in program GS-2017B-FT-12 using the GMOS and FLAMINGOS-2 instruments. GMOS was used to image the field in the 
NIR $Z$-band. Twelve 300s exposures were obtained. 
We used the standard GMOS data reduction package in IRAF to subtract the bias, apply the flat field, and coadd the final images. 
FLAMINGOS-2 imaging was performed through the $J$ and  $K_s$ filters.  
There were 14 $\times$ 60s observations in $J$ band (dithering between each) and 114 $\times$ 8s observations in the $K_s$ filter (dithering every other observation). 
The FLAMINGOS-2 data were dark subtracted, flat-fielded and combined, again using the standard Gemini IRAF package. 



\emph{HST} observations of the field of HE~0515-4414 from program 14594 (P.I.\ R.\ Bielby) using the WFC3 instrument were recently made available in the {\em HST} archive. These consist of broad-band images in the F140W and F160W filters, and grism spectra using the G141 dispersive element. The broad-band images were combined using astrodrizzle, and the aXe software (K\"{u}mmel et al.\ 2009) was used to extract the grism spectra. 
We used the {\em HST} grism spectrum of the quasar to estimate the mass of the black hole, $M_{\bullet}\approx 4.3\times 10^{10}\,M_{\odot}$, based on the width of the H$\beta$ line (9440~km~s$^{-1}$) and the continuum 
flux from 2MASS $J$-band data (Skrutskie et al.\ 2006) using the formulation of Bennert et al.\ (2015).


\section{Results and Discussion}

\subsection{The quasar host galaxy and its environment}
\begin{figure*}
\includegraphics[scale=0.7]{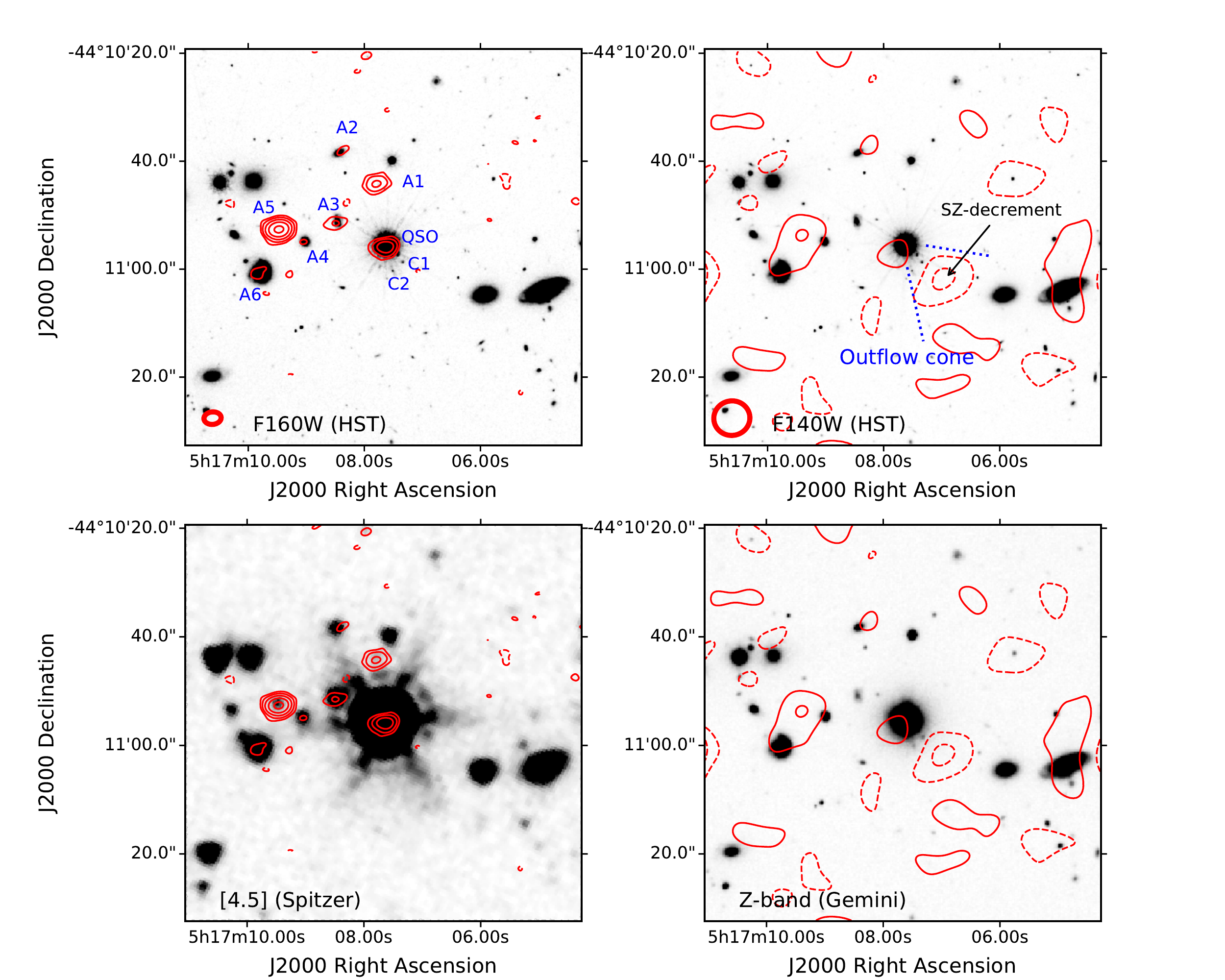}
\caption{Greyscale images of the field of HE\,0515-4414 with ALMA contours superposed in red. The thick red ellipses indicate the ALMA synthesized beam. {\bf Top left:} the {\em HST}/WFC3 F160W data with contours of the naturally-weighted ALMA image. Contours are at -7, 7, 14, 28, 56 and 128$\,\mu$Jy. ALMA sources A1$-$A6, the quasar (QSO), and possible optical/infrared wind components C1 and C2 are labelled. {\bf Top right:} the {\em HST}/WFC3 F140W data with contours of the $uv$-tapered ALMA image. Contours are at  -20, -10, 10, 20$\,\mu$Jy, corresponding to -3, -1.5, 1.5 and 
3$\sigma$. The ALMA contours do not have the primary beam correction applied. This correction is a factor of 10 at the edges of the region shown here, and about a factor of 1.4 halfway between the centre and the edge. The SZE decrement measured from the ALMA data is labelled, and a putative outflow cone encompassing the SZE detection and the optical/infrared wind emission is indicated. 
{\bf Bottom left:} the {\em Spitzer} 4.5$\,\mu$m data with contours of the naturally-weighted ALMA image. {\bf Bottom right:} the Gemini/GMOS $Z$-band image with contours of the $uv$-tapered ALMA image.}

\label{fig:he0515_multiwavelength}
\end{figure*}

The Gemini and {\em HST} images show emission features to the South and West of the quasar in the $Z$, $J$, $F140W$ and $F160W$-bands,  apparently from the host galaxy. 
Their nature is unclear from the imaging alone. However, using the {\em HST} grism 
spectroscopy, we find [O{\sc iii}] emission and faint continuum in the C1 component, which appears to be both extended and 
characterized by a significant velocity gradient ($\approx +1100$ km~s$^{-1}$ over 0\farcs24). This is consistent with a warm component to the 
wind. C1 is 37~kpc from the quasar, and if the gas were traveling at 1000 km~s$^{-1}$, it would take $\approx 4\times 10^7$~yr to reach that distance (depending on projection effects). No line emission was visible from C2.
There is also a discrete emission component between C1 and the quasar, but it is too close to the quasar to obtain a spectrum from the grism data. Taken together with the SZE detection, we hypothesize that C1 and C2 help to define a broad outflow 
cone, as indicated in the top-right panel of Figure~\ref{fig:he0515_multiwavelength}.

The {\em Spitzer}, {\em HST} and Gemini observations can also be used to constrain the environment of the quasar (our {\em Spitzer} data are sensitive to $\sim 10^{11} \, L_{\odot}$ galaxies out to $z\sim 5$ (e.g.\ Falder et al. 2011). Bielby et al.\ (2017) find a poor cluster at $z=0.28$ in the field based on absorption lines in the spectrum of the quasar and follow-up integral field spectroscopy. The estimated mass of this cluster, $6\times 10^{12}M_{\odot}$ is much too small for it to contribute to the observed SZE though.
The lack of any other obvious group or cluster in the field argues against significant contamination of the SZE signal by the intracluster medium of a compact cluster or intragroup medium of a group of galaxies either associated with the quasar, or along the line of sight to the SZE signal. 


\subsection{Interpretation of the SZE Signal}

\label{sec:theory}
We next discuss the interpretation of the SZE signal in the context of simple models. 
The peak brightness of the SZE
decrement of $-$25.9 $\mu$Jy  (Section \ref{sec:alma})
corresponds to a brightness temperature change of $\Delta T_{\rm b,peak} = -37.7$ $\mu$K.
The average projected radius of the bubble is $r \approx 3\farcs9 = 33$ kpc.
The peak is located at a distance of $r_{\rm max} \approx 10\farcs1 = 85$ kpc from the QSO.
The distance from the QSO to the outer edge of the bubble is about $R  \approx 12\farcs8 = 108$ kpc.
(All of these sizes have been corrected for the synthesized radio beam.)

The change in the intensity of the CMB due to the SZE is given by (e.g.\ Sazonov \& Sunyaev 1998):
\begin{equation}
\Delta I(x) = \frac{2 k_{\rm B} T_{\rm CMB}}{\lambda^2} \frac{x^2 e^x}{(e^x-1)^2} \tau \left( \frac{k_{\rm B} T_e}{ m_e c^2} f_1(x) +  \frac{v_r}{c}  \right),
\label{eq:sz}
\end{equation}
where $x \equiv h \nu / ( k_{\rm B} T_{\rm CMB} $); $\tau$ is the Thompson scattering optical depth through the plasma, $\tau = \sigma_T \int dl \, n_e(l)$;  and $f_1(x) = x \, {\rm coth}(x/2)-4$  expresses the frequency dependence of the thermal SZE. 
The $v_r/c$ term is due to the kinematic SZE, which is proportional to the line of sight velocity $v_r$. This expression neglects higher order terms in $v_r/c$ and $k_{\rm B} T_e / m_e c^2$, which are small here.

\subsubsection{Thermal SZE Model}
\label{sec:theory_tSZ}

We first consider a pure thermal SZE model [$v_{\rm r}/c << (k_{\rm B} T_{\rm e})/( m_{\rm e }c^2) f_1(x)$]. For the observed peak decrement, this gives a Compton parameter $y=\tau k_{\rm B} T_{\rm e}/(m_{\rm e}c^2) \approx 2.08  \times 10^{-5}$.
The electron pressure in the bubble is then
$n_e k_{\rm B} T_e = [ y / ( 2r )] ( m_e c^2 / \sigma_T ) \approx 1.26 \times 10^{-11}$~Pa. 
The minimum amount of energy injected into the bubble by the wind is given by the enthalpy
$ (3/2) n k_{\rm B} T V  + p V = (5/2) n k_{\rm B} T V$.  Here, $n \approx 1.92 n_e$ is the total number density of thermal particles, $T$ is an average temperature, and we assume equipartition of ions and electrons, so that $T = T_e$, and $V$
is the volume of the bubble. Under these assumptions, the total energy injected into the bubble by the outflow is:
\begin{equation}
E_{\rm Bub} \approx 2.67 \times 10^{53} \left( \frac{r}{33 \, {\rm kpc} } \right)^2
\left( \frac{y}{2.08 \times 10^{-5} } \right) \, {\rm W} \, .
\label{eqn:EBub}
\end{equation}
This is similar to the enthalpy content of the largest cavities in  
the ICM produced by radio jets (e.g., B\^{i}rzan et al.\ 2008).

In a more realistic scenario,  the expansion of the bubble is supersonic and the kinetic energy of expansion and the shock energy in the IGM need to be added to the enthalpy of the bubble.
We base our treatment of the expansion of the wind bubble on the model in RS11, which uses the self-similar stellar wind bubble solution of Weaver et al.\ (1977).

The RS11 model makes several assumptions.
The first is that cooling of the hot phase gas is negligible.
The second is that the density of the IGM can be estimated from the typical cosmological bias parameter for quasars combined with an estimate of
the over-density of the halo.
Third, the model is spherically symmetric, which our outflow clearly is not.
To remove the assumption of spherical symmetry, we assume that our observed SZE signal comes from a region which is a spherical cone, with the cone apex and the center of curvature located at the QSO.
We estimate the half angle of the cone in our system to be $\theta = 30^{\circ}$ 
(Figure~\ref{fig:he0515_multiwavelength};  upper right). We further assume that the axis of the outflow is at an angle of 45$^{\circ}$ to the outward extension of our line of sight. We assume that the conical bubble expands in the same manner that it would as a portion of a spherically symmetric outflow.
With this simplification, the outer radius of the cone is the same as it would be in a spherically symmetric model with a total wind luminosity of $L_W^* = L_W / f_\Omega$, where $L_W$ is the wind kinetic luminosity of the observed SZE bubble, and $f_\Omega = (1 - \cos \theta )/2 \approx 0.067$ is the fraction of the total 4$\pi$ ster of the outflow. 
Then, equations~(6) \& (7) in RS11 imply that the outer radius of the SZE bubble is:
\begin{equation}
R \approx 45
\left( \frac{L_W}{10^{12} \, L_\odot } \right)^{1/5}
\left( \frac{t}{10^7 \, {\rm yr} }  \right)^{3/5}
\, {\rm kpc} \, ,
\label{eqn:radius}
\end{equation}
where $t$ is the wind lifetime.
The projected outer radius of the observed bubble is 108 kpc, so that the actual radius is about 112 kpc.

Since the pressure within the bubble is expected to be nearly constant (see Figure 1 in RS11),
the peak $y_{\rm max}$ will occur approximately on the longest path length $l_{\rm max}$ through the bubble, which can be easily calculated given the simple geometry.
The maximum $y$ value is then approximately $y_{\rm peak} [ l_{\rm max} / ( 2 R ) ]$,
where $y_{\rm peak}$ is the central peak value in the RS11 model.
Comparison to the $y$ profile in Figure 1 in RS11 shows that the gentle pressure increase in the outer regions of the bubble leads to a correction of 8\%.
This gives
\begin{equation}
y_{\rm max}  \approx 0.916 \times 10^{-5} 
\left( \frac{L_W}{10^{12} \, L_\odot } \right)^{3/5}
\left( \frac{t}{10^7 \, {\rm yr} }  \right)^{-1/5}
\, .
\label{eqn:ymax}
\end{equation}

\begin{table*}
\caption{Properties of conical RS11 models for SZE decrement}
\label{tab:models}
\begin{tabular}{lcccccc}
\hline
Model & Kinetic Power $L_{\rm W}^{\dag}$  & Age $t$ & Energy $E_{\rm W}^{\dag}$ & Shock Speed $v_s$ & Shock Temperature $T_s$ & Ratio {kSZE/tSZE} \\
           & ($10^{11} \, L_\odot$) & ($10^7$ yr) & ($10^{53}$ J) & (km~s$^{-1}$)  & ($10^7$ K) & \\
\hline
tSZE only         & 1.70\phn & \phn8.13 & 1.68 & 807 & 0.91 & 0.95 \\
tSZE plus kSZE &  0.429       &     12.87  & 0.67 & 510 & 0.36 & 1.50 \\
\hline
\end{tabular}

\noindent
$^{\dag}$Assumes an intrinsically one-sided outflow and should be doubled for a two-sided outflow.
\end{table*}           

Equations (\ref{eqn:radius}) and (\ref{eqn:ymax}) are two relations with two
unknowns.
Since we have observed values for both $R  \approx 112$ kpc (corrected for projection) and
$y_{\rm max} \approx 2.08 \times 10^{-5}$,
we can estimate both $L_W$ and $t$, and the total energy of the
event, $E_W = L_W t $.
We can also determine the
shock speed at the outer edge of the bubble as $v_s = (d R / dt ) = (3/5) R / t $.
The velocity in the shocked IGM just within the shock is $v_2 = (3/4) v_s $.
The temperature behind the shock front, $T_s$, is
given by the strong shock jump condition,
$T_s = 3 \mu m_p v_s^2 / (16 k_{\rm B})$. Values for the purely thermal SZE model are given in row 1 of Table~\ref{tab:models}.

\subsubsection{Thermal Plus Kinetic SZE Model}
\label{sec:theory_tkSZ}

In the conical RS11 tSZE wind bubble model presented above,
the post-shock temperature is $T_s \approx 0.9 \times 10^7$ K, and the the post-shock bulk velocity is $v_2 = (3/4) v_s \approx 605$ km~s$^{-1}$. Comparison to Equation \ref{eq:sz} shows that the kSZE term cannot be ignored.
Furthermore, most astrophysical outflows are symmetric, but  in HE 0515$-$4414, there is an SZE decrement to the SW of the QSO, with no corresponding feature to the NE (Figure~\ref{fig:he0515_multiwavelength}; upper right). There are a number of possible explanations, including an intrinsically one-sided outflow, or the counter-flow being blocked by a higher ISM density in its path. However, a combined tSZE and kSZE model provides a 
natural explanation for the one-sided appearance. In the receding half of the outflow (relative to the observer) the kZSE and tSZE contributions to the SZE are both of the same sign, but in the approaching half the two 
partially cancel. Thus, if the contributions of the tSZE and kSZE are similar in magnitude, we would expect to see a one-sided SZE decrement.

We thus consider a model in which the observed SZE decrement is due to significant contributions from both the tSZE and kSZE.
Again, we adopt the conical RS11 wind bubble model. Equation (\ref{eqn:radius}) for the radius remains unchanged.
Adding this kSZE term to the right side of Equation~(\ref{eqn:ymax}) 
yields the maximum total tSZE plus kSZE intensity. The parameters of the tSZE plus kSZE solution are given in the last row of Table~\ref{tab:models}.
The wind luminosity is roughly a factor of 3-4 smaller, and the lifetime is almost twice that for the pure tSZE model, as expected from the dependence of the radius $R$ and $y_{\rm max}$ on $L_W$ and $t$.
In this model, 
about 60\% (40\%) of the SZE decrement is due to the kSZE (tSZE). We note that, in this model, the shock speed is relatively low ($\approx 500\;$kms$^{-1}$), and it is possible that if the quasar is in a sufficiently massive virialized dark matter halo ($\stackrel{>}{_{\sim}}10^{13} M_{\odot}$) with a sound speed comparable to the estimated shock speed, the assumption of a strong shock may not be valid. However, we detect no SZE from the quasar (even after tapering the beam to 20\farcs), suggesting that no such halo is present. Furthermore, the fair agreement in the total energy estimate between the three different methods (total enthalpy, tSZE only and kSZE+tSZE) suggests that our estimate of that quantity at least is fairly robust.

The relatively long lifetime of the wind bubble means that the
possibility of cooling (dominated by bremsstrahlung) needs to be considered.
RS11 give a prescription
for estimating the cooling in their model. The average pressure in the bubble implies $n_e \approx 0.16$ cm$^{-3}$. 
This corresponds to a cooling time of $\sim$0.6~Gyr,  suggesting that cooling effects can be ignored. The equipartition time between protons and electrons is short, $\approx 4000$ yr (e.g., Wong \& Sarazin 2009). Thus, the assumption of equipartition is justified.

Given the relatively low power of the wind, it is also worth investigating
whether it can be produced by a starburst in the quasar
host galaxy.
For the conical outflow model above, assuming a mass loading
factor of unity and a wind velocity of 1000 km s$^{-1}$, a star formation rate of approximately $500 \, M_\odot$ yr$^{-1}$ sustained for $1.3 \times 10^8$ yr
could produce our observed signal.
This would imply the formation
of $\sim$$7 \times 10^{10} \, M_\odot$ of stars over the lifetime of the starburst.
In the absence of constraints on the star formation rate
in this object, we cannot
currently rule out a starburst origin.

HE 0515$-$4414 is
undetected in the cm radio ($<$6\,mJy at 843\,MHz in
the SUMSS survey, Mauch et al. 2003, corresponding to a radio
luminosity at 1.4 GHz, $L_{1.4} <  6 \times 10^{25}$ W Hz$^{-1}$ assuming a spectral index of -0.8), but even a relatively weak radio source can have significant feedback effects (e.g.,
McNamara et al. 2014). Further radio continuum observations are therefore
needed to better constrain any radio AGN activity that might be contributing to feedback.

\subsection{Discussion}

Most theoretical models predict wind kinetic luminosities $\approx 1-10$\% of the bolometric luminosity of the 
quasar. Although the wind luminosity seen in HE0515-4414 is high in absolute terms, it is only a small fraction ($\approx 0.003$\%) of the bolometric luminosity of the quasar ($\approx 1.3\times 10^{15}L_{\odot}$). It is possible that we are seeing HE~0515-4414 in a short-lived extreme outburst  (the quasar is radiating at about the Eddington limit). If this is the case, the mean luminosity over the lifetime of the wind may have been lower. As an alternative to comparing the wind luminosity to the current power of the quasar, we can compare $E_{\rm W}$ to the total radiative energy of the black hole (assuming a radiative efficiency of $\eta=0.1$). This fraction, $E_{\rm W}/(\eta M_{\bullet} c^2)\approx 0.01$\% (0.02\% if the wind is two-sided), is still low compared to the models. 


Based on this simple analysis, we thus obtain an order of magnitude estimate of the relative strength of the quasar wind of $\sim $0.01\% of the averaged quasar luminosity. 
We emphasize though that this is only a single object, and that studies of further objects are needed to confirm this in the general quasar population. One other implication of this study is that although the wind is weak, the cooling time is long. Thermal winds could thus inhibit gas accretion onto the host over longer timescales than is usually assumed for ``quasar mode'' feedback, in much the same way as the non-thermal plasma bubbles in ``radio mode'' feedback. 

\section{Acknowledgments}
This paper makes use of the following ALMA data: ADS/JAO.ALMA\#2016.1.00309.S. ALMA 
is a partnership of ESO (representing its member states), NSF (USA) and NINS (Japan), 
together with NRC (Canada), MOST and ASIAA (Taiwan), and KASI (Republic of Korea), in 
cooperation with the Republic of Chile. The Joint ALMA Observatory is operated by 
ESO, AUI/NRAO and NAOJ. The National Radio Astronomy Observatory is a facility of the 
National Science Foundation operated under cooperative agreement by Associated Universities, Inc.
We thank the North American Data Analysts for their contributions. SC acknowledges support from the department of science and technology, Govt. of India through the SERB Early career research grant. 




\end{document}